\documentclass{article}
\usepackage{emulateapj,amsfonts,psfig}
\begin{document}
\newcommand{\be}{\begin{equation}}
\newcommand{\en}{\end{equation}}
\def\ltsima{$\; \buildrel < \over \sim \;$}
\def\lsim{\lower.5ex\hbox{\ltsima}}
\def\loe{\lower.5ex\hbox{\ltsima}}
\def\gtsima{$\; \buildrel > \over \sim \;$}
\def\gsim{\lower.5ex\hbox{\gtsima}}
\def\goe{\lower.5ex\hbox{\gtsima}}
\def\rref{\par\noindent\hangindent=1.5truecm}
\def\aa #1 #2 {A\&A #1 #2}
\def\aass #1 #2 {A\&AS #1 #2}
\def\araa #1 #2 {ARA\&A #1 #2}
\def\mon #1 #2 {MNRAS #1 #2}
\def\apj #1 #2 {ApJ #1 #2}
\def\apjss #1 #2 {ApJS #1 #2}
\def\apjl #1 #2 {ApJ #1 #2}
\def\astrj #1 #2 {AsJ #1 #2}
\def\nat #1 #2 {Nature #1 #2}
\def\pasj #1 #2 {PASJ #1 #2}
\def\pasp #1 #2 {PASP #1 #2}
\def\msai #1 #2 {Mem. Soc. Astron. Ital. #1 #2}

\def\ass #1 #2 {Ap. Sp. Science #1 #2}
\def\sci #1 #2 {Science #1 #2}
\def\phrevl #1 #2 {Phys. Rev. Lett. #1 #2}
\newcommand{\si}{\left(\frac{\sigma}{0.005}\right)}
\newcommand{\psrdot}{\frac{\dot{P}_{-20}}{P_{-3}^3}}
\newcommand{\ggg}{$\gamma$}
\newcommand{\eee}{$e^{\pm}$}
\newcommand{\ergs}{\rm \ erg \; s^{-1}}
\newcommand{\msol}{\su M_{\odot} }
\newcommand{\etal}{et al.\ }
\newcommand{\Po}{$ P_{orb} \su$}
\newcommand{\pot}{$ \dot{P}_{orb} / P_{orb} \su $}
\newcommand{\myr}{ \su M_{\odot} \su \rm yr^{-1}}
\newcommand{\ppp}{ \dot{P}_{-20} }
\newcommand{\ci}[1]{\cite{#1}}
\newcommand{\bb}[1]{\bibitem{#1}}
\newcommand{\pdot}{ $\dot{P}_{orb}$ \su}
\newcommand{\befl}{ \vspace*{-17pt} \begin{flushright}}
\newcommand{\enfl}{\end{flushright}}
\def\deg {^\circ}
\def\mdot {\dot M}
\def\kms  {\rm \ km \, s^{-1}}
\def\cms  {\rm \ cm \, s^{-1}}
\def\gs   {\rm \ g  \, s^{-1}}
\def\cmtre {\rm \ cm^{-3}}
\def\cmdue {\rm \ cm^{-2}}
\def\gcmdue {\rm \ g \, cm^{-2}}
\def\gcm  {\rm \ g \, cm^{-3}}
\def\rsole {~R_{\odot}}
\def\msole {~M_{\odot}}
\def\fH {{\cal H}}
\def\op {{\cal K}}
\def\nupa{\vfill\eject\noindent}
\def\der#1#2{{d #1 \over d #2}}
\def\inizio{\2acapo\penalty+10000}
\def\fine{\acapo\penalty-10000\blank}
\received{~~} \accepted{~~}
\journalid{}{}
\articleid{}{}

\title{On the bolometric quiescent luminosity and luminosity swing of black
hole candidate and neutron star low mass X--ray transients}

\author{Sergio Campana}

\affil{Osservatorio Astronomico di Brera, Via Bianchi 46, I-23807
Merate (LC), Italy,\\
e-mail: campana@merate.mi.astro.it}

\and

\author{Luigi Stella}

\affil{Osservatorio Astronomico di Roma, Via Frascati 33,
I-00040 Monteporzio Catone (Roma), Italy, \\
e-mail: stella@coma.mporzio.astro.it}

\begin{abstract}
Low mass X--ray transients hosting black hole candidates display
on average a factor of $\sim 100$ larger swing in the minimum
(quiescent) to maximum (outburst) X--ray luminosity than neutron
star systems, despite the fact that the swing in the mass inflow
rate is likely in the same range. Advection dominated accretion
flows, ADAFs, were proposed to interpret such a difference,
because the advected energy disappears beyond the event horizon in
black hole candidates, but must be radiated away in neutron star
systems. The residual optical/UV emission of quiescent low mass
X--ray transients, after subtraction of the companion star
spectrum, was originally ascribed to optically thick emission from
the outer accretion disk regions, where matter accumulates.
Difficulties with this interpretation, led to a revised ADAF
model where the bulk of the residual optical/UV emission in
quiescence does not originate in the outermost disk regions
but is instead produced by synchrotron radiation in the ADAF, and
therefore is part of the ADAF's luminosity budget. We 
demonstrate that, once the residual optical/UV emission is taken 
into account, the bolometric luminosity swing of black hole 
candidates is consistent with that of neutron star systems. 
Therefore ascribing the bulk of the residual optical/UV 
flux to the ADAF removes much of the evidence on which ADAF models 
for low mass X--ray transients were originally developed, namely 
the higher luminosity swing in black holes than in neutron stars.
We also find that, for the neutron star spin periods (a few ms) and
magnetic fields ($\sim 10^8-10^9$~G) inferred from some low mass
X--ray transients, the mass to radiation conversion efficiency of
recently proposed ADAF/propeller models is considerably higher
than required to match the observations, once the contribution
from accretion onto the magnetospheric boundary is taken into
account. Motivated by these findings, we explore here an
alternative scenario to ADAFs in which very little mass accretion
onto the collapsed star (if at all) takes place in the quiescence 
intervals, whereas a sizeable fraction of the mass being transferred 
from the companion star (if not all) accumulates in an outer disk region.
As in some pre-ADAF models, the residual optical/UV emission of 
black hole candidate systems are expected to derive 
from the gravitational energy released by the
matter transferred from the companion star at radii comparable to
the circularisation radius. The quiescent X--ray luminosity 
originates either from accretion onto the black hole candidates at
very low rates and/or from coronal activity in the companion
star or in the outer disk. For comparably small mass inflow rates, 
it can be concluded that the neutron stars in these systems are
likely in the radio pulsar regime. In the interaction of the
radio pulsar relativistic wind with matter transferred from the
companion star, a shock forms, the power law-like emission of
which powers both the harder X--ray emission component and most of the
residual optical/UV observed in quiescence. The soft, thermal-like
X--ray component may arise from the cooling of the neutron
star surface in between outbursts or, perhaps, heating of the
magnetic polar caps by relativistic particles in the radio pulsar
magnetosphere. This scenario matches well both the X--ray and
bolometric luminosity swing of black hole candidate as well as
neutron star systems, for comparable swings of mass inflow rates
toward the collapsed object.
\end{abstract}


\section {Introduction}

Transient X--ray binaries are characterised by a luminosity that
varies over many orders of magnitude, allowing to investigate
accretion onto collapsed stars over a much larger range than
persistent sources. Low mass X--ray transients (LMXRTs), i.e.
transients with a low mass donor star, host either a
sporadically accreting black hole candidate, BHC, or a neutron star,
NS (e.g. Tanaka \& Shibazaki 1996; Campana et al. 1998a).
The outbursts of these transients are likely caused by an
instability of the accretion disk (Cannizzo, Wheeler \& Ghosh 1985;
van Paradijs 1996). The increased matter inflow propagates from the
outer regions of the accretion disk inwards, as testified from the
observation of a delay in the increase of the X--ray
flux relative to optical flux
at the outburst onset of GRO J1655--40 (Hameury et al. 1997) and Aql X-1
(Shahbaz et al. 1998). When determined, the characteristics of the
companion star and binary system are usually similar across
BHC and NS LMXRTs: K dwarf companion
stars and orbital periods in the 2--30~hr range are common
(exceptions are the BHCs GS 2023+338 with an orbital period of
$P_{\rm orb}=155$~hr and GRO J1655--40 with $P_{\rm orb}=62.9$~hr).
Therefore, the time-averaged mass exchange rates in these
systems is expected to be comparable (Menou et al. 1999, hereafter M99).
It is also plausible that the swing between minimum (quiescent)
and maximum (outburst) mass inflow rates towards
the collapsed object is comparable in BHC and
NS LMXRTs.

Observations show that the ratio of minimum X--ray luminosity
in quiescence, $L^X_{\rm min}$, to maximum X--ray luminosity in
outburst, $L^X_{\rm max}$, is significantly smaller (a factor of
about $100$) in BHC than in NS LMXRTs. On average this results
from both a higher $L^X_{\rm max}$ and a lower $L^X_{\rm min}$
in BHC systems (Narayan, Garcia \& McClintock 1997, hereafter N97; Garcia
et al. 1998, hereafter G98; M99).
Eddington-limited accretion is likely responsible for the fact that
only BHCs, being more massive than NSs, achieve $L^X_{\rm max} >
10^{38.5}$ erg s$^{-1}$. N97 (see also G98 and M99)
argue that the smaller values of $L^X_{\rm min}/L^X_{\rm max}$, as well as
$L^X_{\rm min}$, in BHC transients demonstrate that the mass
to radiation conversion efficiency is considerably lower in quiescence,
as expected for advection-dominated accretion flows,
ADAFs, crossing an event horizon.

\begin{table*}[!htb]
\caption{Table 1: Luminosities$^*$ of LMXRTs (see text).}
\centerline{
\begin{tabular}{llllllll}
\hline
Name       &$P_{\rm orb}$&d    & $\log L^X_{\rm min}$&$\log L^{\rm opt}_{\rm
min}$&  $\log L^X_{\rm min}$& $\log L_{\rm max}$&$\log L_{\rm circ}$\\
     & (hr)       &(kpc)&  (erg$\,$s$^{-1}$)  & (erg$\,$s$^{-1}$) & (erg$\,$s$^{-1}$)$^{\diamond}$
&(erg$\,$s$^{-1}$)  &(erg$\,$s$^{-1}$)\\
\hline
GROJ0422+32 & 5.1        & 3.6 & $<31.9$             & 31.7      & $\sim 32.1$
&37.9 (37.8)       & 32.4\\
A 0620--00  & 7.8        & 1.2 &\ 31.0               & 32.0      & \ 32.0
& 38.4 (38.1)       & 32.3\\
GS 2000+25  & 8.3        & 2.7 & $<32.3$             & 32.6      & $\sim 32.8$
& 38.4 (38.3)       & 32.4\\
GS 1124--684& 10.4       & 6.5 & $<32.6$             & 32.3      & $\sim 32.8$
& 39.1 (38.9)       & 32.1\\
H 1705--250 & 16.8       & 8.6 & $<33.7$             & 34.4      & $\sim 34.5$
& 38.3 (39.6)       & 32.5\\
4U 1543--47 & 27.0       & 8.0 & $<33.7$             & 34.6      & $\sim 34.7$
& 39.6 (39.0)       & 33.2\\
GROJ1655--40& 62.9       & 3.2 &\ 32.4               & 33.6      & \ 33.6
& 38.5 (38.1)       & 33.5\\
GS 2023+338 & 155.3      & 3.5 &\ 33.2               & 34.3      & \ 34.3
& 39.3 (39.3)       & 33.3\\
\hline
SAX J1808.4--3658& 2.0   & 4.0 &\ 32.5               & $<32.6$   & $\sim 32.5$
& 36.8 (---)        & 33.0\\
4U 2129+47  & 5.2        & 6.3 &\ 32.7               & $\sim 30$ & \ 32.7
& 38.2 (---)        & 31.9\\
Cen X-4     & 15.1       & 1.2 &\ 32.7               & 32.2      & \ 32.8
& 38.1 (38.0)       & 32.0\\
Aql X-1     & 18.9       & 2.5 &\ 32.8               &$\lsim 32$ & $\sim 32.8$
& 37.6 (37.6)       & 32.0\\
4U~1608--52 & ---        & 3.3 &\ 33.3               &$\lsim32.3$& \ 33.3
& 38.0 (37.7)       & --- \\
X~1732--304 & ---        & 4.5 &\ 33.1$^{\dag}$      & ---       & \ 33.1
& 37.8 (---)        & --- \\
EXO 0748--676& 3.8       & 3.8 &\ 34.0               & $\sim 30$ & \ 34.0 
& 37.5 (37.3)       & 32.4\\
\hline
\end{tabular}
}

\noindent $^*$ Distances are from G98 (see also the text).
$L^X_{\rm min}$ is the quiescent $0.5-10$~keV luminosity, $L^{\rm
opt}_{\rm min}$ the quiescent optical/UV luminosity after
subtraction of the contribution from the companion star and
$L^{\rm bol}_{\rm min}$ the quiescent {\it bolometric} luminosity
as estimated from the sum of $L^X_{\rm min}$ and $L^{\rm opt}_{\rm
min}$. $L_{\rm max}$ is the maximum luminosity, as inferred from
the $0.5-10$~keV luminosity at the outburst peak (see G98); values
in parentheses are from Chen, Shrader \& Livio (1997; hereafter
C97). $L_{\rm circ}$ is the energy released by accretion at the
circularisation radius, according to standard mass transfer
models.

\noindent $^{\diamond}$ In those cases in which the upper limits on the X--ray
luminosity is larger or comparable to the measured residual optical/UV
luminosity after subtraction of the contribution from the companion star,
we have adopted the latter value for the bolometric luminosity.

\noindent $^{\dag}$ 2--10 keV luminosity.

\end{table*}

In this context, the residual optical/UV luminosity of quiescent systems,
after subtraction of the contribution from the mass donor star, has been
interpreted in different ways. In the original ADAF models of Narayan,
McClintock \& Yi (1996), the residual optical emission is supposed to
originate from the outermost disk regions, where the optically thick
standard model applies.
Emission from the hot spot where the accretion stream from the mass
donor impacts the disk likely contributes to the optical flux.
Motivated by Wheeler's (1996) argument that the effective
temperature of such an outer accretion disk would be in the unstable regime
(see also Lasota, Narayan \& Yi 1996),
Narayan, Barret \& McClintock (1997) revised the original model and
proposed that the residual optical/UV luminosity derives from synchrotron
emission in the ADAF, while the outer standard disk region is cooler and
farther from the BH, and emits mainly in the infrared (see Menou, Narayan
and Lasota 1999). It should be emphasised that, in this
case, besides the X--ray emission, the residual optical/UV
emission should therefore be included in the energy budget of the ADAF.
In Section 2 we review the X--ray and optical properties of BHCs and NSs
and compare their minimum to maximum luminosity ratios with and without the
inclusion of the residual optical/UV quiescent luminosity.
As a result no clear 
distinction is found between the luminosity ratio of these 
two classes if the residual optical/UV luminosity is included in the budget.

\begin{figure*}[!h]
\psfig{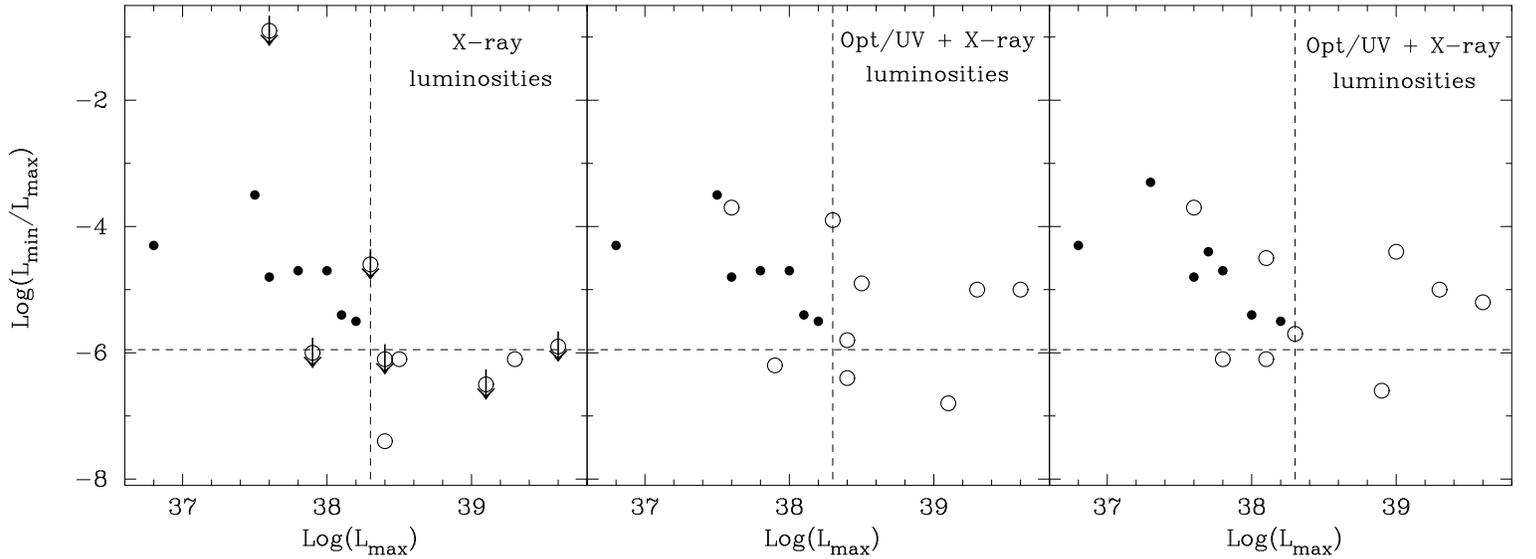}
\caption{Minimum to maximum
luminosity ratios versus maximum luminosity of NS (filled circles)
and BHC (open circles) LMXRTs. The left panel shows the results
based on the X--ray data alone ($L^X_{\rm min}/L_{\rm max}$).
Arrows indicate upper limits. Maximum luminosities are according to G98.
In the middle panel the luminosity ratio $L^{\rm bol}_{\rm min}/L_{\rm max}$
includes the contribution of optical/UV photons (after subtraction the
contribution of the companion star). The right panel is the same as the
middle panel, except that the maximum luminosities estimated by C97
are used.
Dashed lines represent the values of $L^X_{\rm min}/L_{\rm max}$
and $L_{\rm max}$ proposed by N97 to separate BHC from NS LMXRTs.}
\label{figdue}
\end{figure*}

In Section 3 we comment on some difficulties in the
application of ADAF scenarios to quiescent BHC and NS systems.
We then consider an alternative
scenario for the quiescent emission of LMXRTs, which does not involve
an ADAF solution.
The basic ansatz of this scenario is that a very small fraction of the
transferred matter (if at all) reaches the compact object during quiescence; 
rather most of the transferred matter is supposed to 
accumulate close to the circularisation radius or lost in a wind
(Section 4).
As in pre-ADAF models of BHC LMXRTs, the implied ratio of quiescence to 
outburst mass inflow rates towards the collapsed star is 
$10^{-7}-10^{-8}$ (or smaller). We show that, if the same ratio applies 
also to the fast spinning weakly magnetic NSs hosted in LMXRTs, then 
the radio pulsar regime is expected to operate in the quiescent state, 
with shock emission driven by the pulsar wind dominating 
the quiescent luminosity and spectrum of these systems. In section 5  
some properties of quiescent LMXRTs are also compared with the expectations
of our model. Our conclusions are summarised in Section 6.

\section{Bolometric luminosities of quiescent Low Mass X--ray Transients}

We estimate here the quiescent optical/UV luminosity of LMXRTs once the
contribution from the mass donor star is subtracted (see also Menou, Narayan
\& Lasota 1999). We adopt the source sample and
X--ray luminosities (0.5--10 keV range) of G98, complemented with recent
results on the NS systems SAX J1808.4--3658
(Stella et al. 2000) and X~1732--304 (Guainazzi et al. 1999). Our results
remain unchanged if the  X--ray luminosities derived by
C97) are used  in place of those in G98 (see also
Tab. 1 and Fig. 1). 
Being many orders of magnitude higher than
the optical luminosity, the maximum X--ray luminosity
provides a reliable estimate
of the bolometric luminosity at the outburst peak
(i.e. $L_{\rm max}^X=L_{\rm max}^{\rm bol}=L_{\rm max}$).
Optical V magnitudes and  absorptions are from C97,
unless otherwise specified.

\subsection{Black hole candidates}

The BHC transient A~0620--00 has a quiescent 0.5--10 keV luminosity
of $L^X_{\rm min}\sim10^{31}\ergs$ (for a distance of 1.2 kpc). This
value is obtained by extrapolating the ROSAT data and using a fixed column
density of $N_H=1.2\times10^{21}\cmdue$ (McClintock et al. 1995).
Due to the small number of collected photons in the ROSAT observation
($\sim 40$) the spectrum is very poorly determined and can be well fit by
a variety of single component models.
A short wavelength HST/FOS spectrum of the quiescent optical counterpart
yielded a 1350--2200\,\AA\ luminosity of $0.6-4\times 10^{31}\ergs$.
These results have been confirmed by higher quality HST/STIS spectra
(McClintock \& Remillard 2000).
At optical wavelengths (2200--4750\,\AA) the spectrum can be fit by a 9000 K
black body (luminosity of $\sim 10^{32}\ergs$), after subtraction of the
$58\pm4\%$ contribution from the K5V star companion, which affects mainly
the spectrum at wavelengths $\gsim 4000$\,\AA.
At longer wavelengths (2.0--2.5\,$\mu$m) the K dwarf flux dominates,
making up $75\pm17\%$ of the infrared luminosity (Shahbaz et al. 1999).
Therefore, the quiescent optical/UV luminosity of A~0620--00
outshines the X--ray luminosity by a factor of $\sim 10$
(see also McClintock et al. 1995). We estimate a rough
bolometric quiescent luminosity of $L^{\rm bol}_{\rm min} \sim
10^{32}\ergs$.

The lowest quiescent X--ray luminosity detected from GS 2023+338 (V~404 Cyg)
is $L^X_{\rm min}\sim 2\times 10^{33}\ergs$ (Narayan, Barret \& McClintock 
1997; Campana 2000). The spectrum is well fit by either a power law
(photon index $\Gamma\sim 1.5-2$) or a bremsstrahlung ($k\,T_{\rm br}\sim 5-10$
keV). GS 2023+338 is significantly reddened ($A_{\rm V} \sim 4$ mag) and no UV
data are available. Casares et al. (1993) estimate that the contribution
of the accretion disk to the optical flux relative to the G9V--K0III
companion is 72\%, 36\% and 19\% in the B, V and R bands,
respectively. These fractions convert to dereddened luminosities
of $L_{\rm B}\sim 8\times 10^{33}\ergs$,
$L_{\rm V}\sim 8\times 10^{33}\ergs$ and $L_{\rm
R}\sim 6\times 10^{33}\ergs$. For GS 2023+338 we therefore adopt a value
of $L^{\rm bol}_{\rm min} \sim 2\times 10^{34}\ergs$, which is a factor of
$\sim 10$ higher than $L_{\rm min}^X$.

GRO J1655--40 has been detected in quiescence at a level of
$L^X_{\rm min} \sim 2\times 10^{32}\ergs$ (Hameury et al. 1997).
The spectrum can be described by a power law model ($\Gamma\sim 1.5$).
Soft X--ray and UV flux measurement are severely hampered by
a large absorption ($A_{\rm V}\sim 4$ mag).
The F5IV companion star outshines the disk in the optical ($95\pm2\%$
at 5500\,\AA; Orosz \& Bailyn 1997).
An estimate of the V-band luminosity of the accretion disk
based on the model by Orosz \& Bailyn (1997) yields
$L_{\rm V} \sim 4\times10^{33}\ergs$.
The optical quiescent luminosity therefore dominates the X--ray emission
by a factor of $\sim 10$.

For all other known BHC transients there are only upper limits to
their quiescent X--ray flux (see Tab. 1). Yet, their optical counterparts
are relatively well studied and the residual optical flux
can be estimated. In particular, Keck spectra were used to estimate the
fraction of the 6600--6800\,\AA\ luminosity that originates from the
companion star. This is 85\%, 95\%, 30\% and 60\% in GS 1124--684
(Nova Mus 91), GS 2000+25 (Nova Vul 88), H~1705--250 (Nova Oph 77) and
GRO J0422+32 (Nova Per 92), respectively (Casares et al. 1997; Harlaftis
et al. 1996, 1997, 1999). All these BHCs have K--M dwarf
companions. V-magnitudes are ${\rm V}=20.5,\ 21.2,\ 21.3$ and 22.2 mag,
respectively. By assuming that the same fractions above hold for the V band,
a very conservative assumption for K--M dwarf stars,
we infer that the dereddened V luminosities of the accretion disk are
$L_{\rm V}=2\times 10^{32},\ 4\times 10^{32},\
2\times 10^{34}$ and $5\times 10^{31}\ergs$, respectively.
The optical counterpart of 4U~1543--47, an A2V star, has V=16.7.
Orosz et al. (1998) estimate a disk contribution of
10\%, 21\% , 32\% and 39\%, in the B, V, R and
I bands, respectively. The increasing disk contribution for longer optical
wavelengths is due to the relatively hot companion.
The dereddened disk V and I luminosities are
$L_{\rm V}\sim 3\times 10^{34}\ergs$ and $L_{\rm I}\sim 10^{34}\ergs$,
respectively. Consequently we estimate $L^{\rm bol}_{\rm min} \sim
4\times 10^{34}\ergs$.

Therefore, we conclude that the bolometric luminosity of
quiescent BHC LMXRTs is dominated by optical/UV disk emission.
When both the X--ray and optical quiescent luminosities are available,
the latter are systematically higher by about an order of magnitude.

\subsection{Neutron stars}

The two best studied NS LMXRTs, Aql X-1 and Cen X-4,
have $L^X_{\rm min} \sim 6\times 10^{32}\ergs$ (Campana et al. 1998b)
and $L^X_{\rm min}\sim 5\times 10^{32}\ergs$
(Asai et al. 1998; Campana et al.
2000), respectively. Recent optical studies of the field of
Aql X-1 indicate that the true optical counterpart is
located $0.5''$ from the previously known star (e.g. Shahbaz, Casares \&
Charles 1997). The magnitude of the counterpart is ${\rm V}=21.6$ mag
(Chevalier et al. 1999).
The dereddened ($A_{\rm V}=1.2$ mag) V luminosity can therefore be
$L_{\rm V}\sim 10^{32}\ergs$ at the most (i.e. if the entire
V luminosity came from the disk). In any case
this luminosity represents a small fraction of the quiescent X--ray
luminosity.

In the case of Cen X-4 the residual optical flux is estimated to contribute
80\%, 30\%, 25\% and 10\% in the B, V (18.7 mag), R and I bands,
respectively (Shahbaz,  Naylor \& Charles 1993).
The corresponding reddening-corrected luminosities ($A_{\rm V}=0.3$
mag) are $L_{\rm B}\sim 9\times10^{31}\ergs$, $L_{\rm V}\sim 5\times
10^{31}\ergs$, $L_{\rm R}\sim 6\times10^{31}\ergs$ and  $L_{\rm I}\sim
2\times 10^{31}\ergs$.  Recently, an UV spectrum has been
obtained with the HST/STIS (McClintock \& Remillard 2000). The main result
is that in a $\nu\,F_{\nu}$ vs. $\nu$ representation there is only factor of
$\sim 1-2$ increase from the X--rays to the optical.
Therefore, the quiescent X--ray and optical luminosities are comparable in
Cen X-4.

The quiescent X--ray state of these two sources has been studied in some
detail with BeppoSAX and ASCA pointed observations.
Their 0.1--10 keV spectrum comprises a soft component,
modeled by a black body with $k\,T_{\rm bb}\sim 0.1-0.3$ keV and equivalent
radius of $\sim 1-3$ km,
plus a hard power law component with photon index $\sim 1-2$
(Campana et al. 1998b, 2000; Asai et al. 1996, 1998).
The contribution of the two spectral components to the
0.5--10 keV luminosity is comparable.

The quiescent X--ray flux of 4U~1608--522, 4U 2129+47 and
EXO 0748--676 has also been detected (e.g. Campana et al. 1998a, see Tab. 1).
4U~1608--522 has been revealed in quiescence at a level  of
$L^X_{\rm min} \sim 2\times 10^{33}\ergs$ (0.5--10 keV for $d=3.3$~kpc;
Asai et al. 1998). The highly absorbed ($A_{\rm V}=5.2$ mag)
optical counterpart of 4U~1608--522 has ${\rm J}=18.0$ mag (${\rm R}>22$ mag)
and a luminosity of $\sim 2\times 10^{32}\ergs$ at
the most (including the companion star).
4U 2129+47 has $L^X_{\rm min}\sim 6\times 10^{32}\ergs$ (0.5--10 keV for
$d=6.3$~kpc). The F9 subgiant companion dominates the optical flux
(V=18.5 mag). Garcia \& Callanan (1999) estimate ${\rm V}=24.5$ mag for the
disk of 4U 2129+47, implying a dereddened ($A_{\rm V}\sim 1.5$ mag) V
luminosity of only $\lsim 10^{30}\ergs$.
EXO 0748--676 has a relatively high quiescent luminosity of  $L^X_{\rm
min}\sim 10^{34}\ergs$ (0.5--10 keV for $d=3.8$~kpc). However, being
a high inclination system, EXO 0748--676
should be treated with caution since its
X--ray flux variations might be driven by geometrical effects (e.g.
obscuration by a variable height of the disk rim) rather than genuine mass
inflow rate variations.  Optical observations provided an upper limit on the
V-magnitude of  the quiescent optical counterpart of EXO 0748--676 (${\rm
V}>23$ mag for  $A_{\rm V}=1.2$ mag); this translates to a V luminosity of
$\sim 10^{30}\ergs$.

In addition to the sources in the G98 sample, SAX J1808.4--3658 (Stella et
al. 2000) and X 1732--304 (Guainazzi et al. 1999) have also been detected
in quiescence.
SAX J1808.4--3658 has a quiescent X--ray luminosity of $2-3\times
10^{32}\ergs$.
The optical counterpart was detected only during the outburst decay.
An upper limit on the quiescent V magnitude of $>20.5$ mag has been derived
(Giles et al. 1999).
By using the galactic column density to estimate
$A_{\rm V}$, we derive an upper limit to the V luminosity of $4\times
10^{32}\ergs$. X 1732--304 in the globular cluster Terzan 1 was previously
considered a persistent (though highly variable) source.
In April 1999 it was observed in a quiescent state at a 2--10 keV
luminosity of $1.4\times 10^{33}\ergs$. The quiescent X--ray spectrum
was compatible with the two-component spectrum inferred for Aql~X-1
and Cen~X-4. The optical counterpart is not known.

All the data above indicate that in quiescent NS LMXRTs
the X--ray luminosity exceeds (or, at the most, is comparable to)
the optical luminosity.

\subsection{Comparison of luminosity ratios and quiescent luminosities}

The ratio of minimum to maximum luminosity of LMXRTs,
as estimated by N97 and G98 on the basis of the X--ray data alone,
is plotted in the left panel of Fig.~1, versus the maximum luminosity.
BHCs are clearly separated from NSs both in terms of maximum luminosity and
X--ray luminosity ratio. The middle panel shows instead
$L^{\rm bol}_{\rm min}/L_{\rm max}$ as estimated above based
on both X--ray and optical measurements (see also Tab. 1): the distinction
between BHC and NS systems is no longer apparent.
Using a Kolmogorov-Smirnov, KS, test we estimate that the values of
$L^{\rm bol}_{\rm min}/L_{\rm max}$ for the two classes of
transients have a 9\% probability of being drawn by chance from the
same parent distribution (note that using the X--ray data alone the
KS probability is $<0.2\%$).
To check that this conclusion is robust and independent of the
method for estimating $L_{\rm max}$ adopted by G98,
we calculated also the minimum to maximum luminosity ratios by using the
$L_{\rm max}$ values derived by C97 (see also Tab. 1).
The results are shown in the right panel of Fig.~1.
Also in this case, it is not possible to distinguish BHC from NS
systems by using $L^{\rm bol}_{\rm min}/L_{\rm max}$
(KS probability of 22\%).
We note that the conclusions above are even strengthened
if EXO~0748-676, and/or SAX J1808.4--3658 and
X 1732--304 (i.e. the sources not included in the G98 sample) are excluded
from the source sample.

Considering minimum bolometric luminosities only, one has that the BHC and
NS transients population have a 56\% probability of being drawn by chance from
the same parent distribution. This probability is 22\% if minimum
bolometric luminosities in Eddington units are used instead.
This is at variance with the results obtained by M99, who used
minimum X--ray luminosities.

These results show that once the contribution from the optical
luminosity (after subtraction of the mass donor's spectrum) is included in
the evaluation of the quiescent luminosity of LMXRTs, there is no evidence
that the luminosity swing of BHCs is larger than that of
NSs, neither that the minimum (quiescent) luminosity
of BHCs is lower. In fact the optical luminosity, while usually
negligible in NS LMXRTs, dominates the quiescent luminosity of BHC
transients.

\section{Comments on ADAF models for Low Mass X--ray Transients}

\subsection{Black hole candidates}

It has long been realised that, if steady accretion takes place in
quiescence, standard optically thick
accretion disk models (with radiative efficiency $\sim 0.1$) are inadequate
to explain both the X--ray luminosity and the temperature vs.
effective area combination inferred from the residual
optical emission of BHC LMXRTs (see McClintock et al. 1995).
While the former problem could be cured by invoking, e.g.
the presence of a hot phase or of a Comptonising inner accretion disk
region (see e.g. Spruit 2000), the latter problem is difficult to solve: if the optically
thick accretion disk extended to (or close to) the marginally stable orbit,
then the area where most of the quiescent flux is emitted would be small
and the corresponding black body temperature ($T_{\rm bb}\sim 10^5 -10^6$ K),
at least, an order of magnitude higher than inferred from the residual
optical emission of quiescent BHC transients (e.g. $T\sim 9000$ K
in A~0620--00, see McClintock et al. 1995).

A considerable amount of work has been carried out in recent years
on ADAF models, where the radiative efficiency is very low ($\sim
10^{-4}-10^{-3}$) and most of the gravitational energy of the inflowing
matter is stored as thermal and/or bulk kinetic energy and advected towards
the collapsed star.  Solutions of this type exist for sub-Eddington mass
accretion rates  ($\mdot < 0.1-0.01\ \mdot_{\rm Edd}$).
The luminosity of an ADAF scales approximately as $\dot M^2$ (as opposed
to the  $\dot M$ scaling of standard accretion). If the accreting
object is a BH, the advected energy disappears beyond the event
horizon. In the case of a NS,
the energy of the ADAF is instead radiated away when the plasma
reaches the star surface.
This implies that, for a given swing of mass inflow rate between
outburst and quiescence, the corresponding swing of accretion luminosity
of BHC transients should be several orders of magnitudes larger than that
of NS LMXRTs (N97).
Even though the observed X--ray luminosity swing of BHCs is only a factor of
$\sim 100$ larger than that of NS systems (whereas simple ADAF models would
predict a factor of $\sim 10^3-10^4$) LMXRTs appeared
to confirm this fundamental property of BHs accreting through an ADAF.
This conclusion hinged upon the ansatz that the minimum X--ray luminosity
provides a reasonable estimate of the quiescent bolometric luminosity
generated by the ADAF (see N97).
Indeed the ADAF model originally suggested by Narayan et al. (1996) to
interpret the spectral energy distribution of the BHC transient A 0620--00
envisaged an
outer optically thick standard disk (radius of $\sim 10^9$~cm), responsible
for the bulk of the emitted optical/UV radiation,
together with an inner ADAF giving rise to the X--ray luminosity.
However, Wheeler (1996, see also Lasota, Narayan \& Yi 1997) pointed
out that the inferred effective temperature of the outer
disk regions is within the unstable range ($\sim 5000-8000$~K).
Moreover, it is difficult to match
the advection region with the outer disk in terms of surface density
and angular momentum.

A revised ADAF model was introduced by Narayan, Barret \& McClintock
(1997) in order to interpret the multiwavelength quiescent spectra of
GS~2023+338 and eliminate the above mentioned difficulties.
This model ascribes the vast majority of the optical/UV
luminosity to synchrotron emission from the ADAF, which
extends from a transition radius of $\sim 10^{10}-10^{11}$~cm inwards.
The outer standard disk is at larger radii
and emits predominantly in the infrared. The bolometric radiative
efficiency of ADAF model constructed in this way is $\sim 10^{-3}$
(as opposed to $\sim 0.1$ in standard accretion), with an X--ray radiative
efficiency a couple of orders of magnitude lower.
One of the consequences of this ADAF model is that the
optical/UV emission, being produced by the ADAF,
is part of the luminosity budget of the ADAF itself
and should therefore be included in any comparison involving the
minimum luminosity of BH and NS LMXRTs. However,
the results in Section 2 show that once the contribution from the optical/UV
luminosity (after subtraction of the mass donor's spectrum) is included in
the evaluation of the quiescent luminosity of LMXRTs, there is no evidence
that the luminosity swing of BHCs is larger than that of
NSs, neither that the minimum luminosity of BHCs is lower.

\subsection{Neutron stars}

Recent ADAF modeling of quiescent NS LMXRTs exploited the idea that most of
the inflowing matter is prevented from reaching the NS surface by the
action of the magnetospheric centrifugal barrier and that the propeller
effect,
ejecting matter from the magnetospheric boundary to infinity, might be at
work (Zhang, Yu \& Zhang 1998; M99). This was in part motivated
by the growing evidence that many low mass X--ray Binaries, LMXRBs, (and
likewise LMXRTs) host fast spinning NS (periods of a few ms) with weak
magnetic fields ($\sim 10^8-10^9$~G, see Section 4.2).
The fact that the observed X--ray luminosity swing of NS LMXRTs is only a
factor of $\sim 100$ smaller than that of BHC LMXRTs played also a role.
An additional issue with simple ADAF models is that,
while the mass accretion rates deduced in BHC LMXRTs from ADAF
modeling of their quiescent X--ray fluxes represent a large fraction
($\sim 1/3$) of the rates predicted by standard mass transfer models
in binaries, the corresponding fraction in NS LMXRTs is very low
($\sim 0.1$\%; M99). This suggested
that only a very small fraction of the quiescent mass inflow rate reaches
the NS surface in order to power the soft (thermal-like)
X--ray emission observed from several quiescent NS LMXRTs.
M99 investigated in detail this possibility and concluded
that an efficient propeller effect, possibly in association with
mass loss in a disk wind, would be required to match the observed
quiescent luminosities of NS LMXRTs.

Independent of whether the matter inflowing onto the magnetospheric
boundary accumulates locally or is ejected to infinity (still an open
problem),
the trouble with the propeller interpretation is that there is no ``sink"
where the energy stored in the ADAF  down to the magnetospheric
radius can be permanently hidden. For the NS parameters deduced for LMXRTs,
the gravitational energy, $L(r_{\rm m})$,
of the accretion flow down to the magnetospheric boundary,
$r_{\rm m}$, (which must be radiated away) is large,
because $r_{\rm m}$ is $\sim 10\, R$ at the most, with $R$
is the NS radius\footnote{$r_{\rm m}$ must be
smaller than the light cylinder radius.}
(Stella et al. 1994; Campana et al. 1998a).
If only a fraction of $f_{\rm R} \sim 10^{-3}$ of the mass
transfer rate from the companion star reaches
the NS surface (M99),
then $L(r_{\rm m})$ would exceed
the luminosity released at the NS  surface, $L(R)$, by a large factor
($R/(r_{\rm m}\,f_{\rm R}) \sim 100$).  The conclusion above remains
basically
unchanged even if a substantial mass  loss took place in the form of a disk
wind. For example in the solution with  the highest disk mass loss discussed
by M99 (transition  radius of $\sim 10^{10}$ cm, and $\mdot(r) \propto
r^{0.8}$, i.e. their ``strong wind'' model), $\sim 1$\% of the mass inflow rate
reaches the magnetospheric boundary implying that $L(r_{\rm m})$ exceeds
$L(R)$ by a factor of $\sim 30$. (This factor would be higher still for disk
models with smaller mass loss.)
In principle, a possible way out could be that
$L(r_{\rm m})$ is released away from the X--ray band at energies that are
not accessible to current instrumentation.
The luminosity released by the ADAF at $r_{\rm m}$
(some $10^{34}-10^{35}\ergs$) together with the size of the emitting region
(roughly $r^2_{\rm m} \sim 10^{14}$~cm$^2$),
would imply a minimum emission temperature of $\sim 0.1-0.2$~keV.
Therefore this luminosity could not be hidden in the EUV; emission at
$\gamma$-ray energies would instead be required in order
to avoid detection by current instrumentation.
We regard this as contrived and conclude that current ADAF/propeller models
for NS LMXRTs do not appear to produce a sufficiently marked reduction of
the accretion to radiation conversion efficiency to match the
observations.

\section{An Alternative Scenario for Quiescent
Low Mass X--ray Transients}

As noted by a number of authors, a considerable luminosity is
produced in quiescent LMXRTs by the release of gravitational
energy, as the stream of matter from the Lagrangian point of the
mass donor star reaches the outer regions of the disk around the
collapsed star. It is easy to see that this luminosity is of the order
of the residual optical/UV luminosity of LMXRTs. The expected 
values of the luminosity, $L_{\rm circ}$, released by the mass transfer
rate at the circularisation radius, $r_{\rm circ}$, (see Lubow \& Shu 1975)
are given in Table~1. The mass transfer rate, $\mdot_{\rm tr}$,
of individual LMXRTs was estimated from binary evolutionary models
(Pylyser \& Savonije 1988; King, Kolb \& Burderi 1996), with
the star masses and orbital period taken from Menou,
Narayan \& Lasota (1999).
If this luminosity is released in the optically thick regime,
equivalent black body temperatures of 6000--16000 K are expected.
These temperatures and the values of $L_{\rm circ}$ are in the range
measured for the residual optical/UV luminosity of LMXRTs. We emphasise 
that such an interpretation would hold even if an ADAF were present
in the inner disk, provided the ADAF itself {\it does not} produce 
the bulk of the optical/UV emission; this is the case for the original 
ADAF model of Narayan et al. (1996; see Section 3).

Whether the emission at $\sim r_{\rm circ}$, is 
azimuthally confined to the 
hot spot where the accretion stream from the mass donor impacts the outer
disk was investigated by McClintock et al. (1983). These authors concluded 
that the modulation in the residual optical/UV flux of A 0620--00 may be 
consistent with that expected from the hot spot. 
Doppler maps (Marsh \& Horne 1988) have been obtained for a number of
quiescent BHC systems  mainly by using the H$\alpha$ emission line
and provide a further test for the geometry
and kinematics of the accreting matter
(Marsh, Robinson \& Wood 1994; Casares et al. 1997; Harlaftis et al.
1996, 1997, 1999).
These maps show the typical ring-like distribution of the
outer regions of an accretion disk, similar to that observed in cataclysmic
variables.
The H$\alpha$ emission is strongest at a location associated with the
interaction between the gas stream and the accretion disk, but extends
also to a factor of $\sim 2$ higher velocity regions ({\it i.e.} 
somewhat smaller radii; e.g. Marsh et al. 1994).

The rate at which mass accumulates in the outer disk during quiescent
intervals, $\dot M_{\rm acc}$, has been estimated from the outburst
recurrence
time and fluence of those LMXRTs which have displayed more than one outburst
(M99). Though uncertain, the
inferred mass accumulation rate is of the order of the mass transfer
rate from the companion star predicted by binary evolutionary models,
$\dot M_{\rm tr}$ (in the context of ADAF models M99
adopt a value of $\mdot_{\rm acc} \sim 1/3\, \mdot_{\rm tr}$).

In consideration of the discussion in Section 3, we resolved to explore an
alternative scenario for quiescent LMXRTs which does not involve an ADAF.
Our basic ansatz is that during quiescence 
only a very small fraction of $\mdot_{\rm tr}$ 
(if at all) accretes onto the collapsed star, while a large fraction of
$\mdot_{\rm tr}$ accumulates in the outer disk or
is lost in a wind (e.g. Meyer-Hofmeister \& Meyer 1999; Blandford \&
Begelman 1999).
We are aware that such a sharp cutoff of the mass inflow rate
from the outer disk regions is not envisaged by current instability models
for LMXRTs. 
Yet, our suggested scenario is reminiscent of earlier (pre-ADAF)
disk instability
models that were proposed  to explain the long recurrence time of 
A 0620--00 (Huang \& Wheeler 1989; Mineshige \& Wheeler 1989). These models
are characterised by a very low
mass transfer rate during quiescence (some $\sim 8$ orders of magnitude
lower then at the outburst peak)
and are capable of reproducing the $\sim 50$ yr outburst recurrence with
an {\it ad hoc}, factor of $\sim 20$, variation of the $\alpha$
viscosity parameter across the upper and lower branches of the surface
density vs. viscosity relationship.
We note that a similar scenario, namely a nearly empty inner
accretion disk in quiescence, has been proposed to explain the delay
between the rise of the optical/UV flux in cataclysmic variables
within the disk instability models (e.g. King 1997).
We also maintain that the swing
between minimum and maximum mass inflow rates is similar
across BHC and NS systems.

Our goal is to show that, if the mass
inflow rate toward the compact object does vary by at least 7--8
orders of magnitude across the outburst/quiescence transition, then
the properties of quiescent BHC and NS LMXRTs can be readily
interpreted. This is done in the next two subsections.

\subsection{Quiescent Black Hole Low Mass X--ray Transients}

If the residual optical/UV luminosity of quiescent BHCs derives
from the release of gravitational energy in the outermost disk
regions by the material transferred from the companion star, then
the faint quiescent X--ray luminosity of these systems might
originate from very low level accretion onto the BHC, or, perhaps,
from intense coronal activity in the outer disk region where matter
accumulates.

For a standard radiative efficiency of $\sim 0.1$, the observed
X--ray luminosities would require that accretion into the BH
takes place at a level of
$\dot M_{\rm acc} \sim 10^{11}-10^{12}$ g s$^{-1}$ during
quiescence. This value is $7-8$ orders of magnitude lower than the
mass accretion rate at the outburst peak and 2--4 orders of magnitudes
lower than $\mdot_{\rm tr}$, implying that only a minute fraction of
the matter reaches the BHC in quiescence.

According to standard optically thick disk models, the emitted spectrum for
a low mass inflow rate ($\lsim 10^{14}\gs$) onto a BH peaks at energies
below 0.2 keV
(e.g. de Kool 1988). However, if the density of disk becomes too
low, thermal equilibrium can no longer be maintained and the accreting
gas heats up giving rise to a hot phase in its innermost regions
(de Kool \& Wickramasinghe 1999).
Such a hot inner corona may be responsible for the low level X--ray
emission.
In this interpretation the quiescent X--ray
luminosity would be produced deep in the potential well ($\sim
10^7$ cm), i.e. a region $\sim 10^{4}$ smaller than $R_{\rm circ}$.
Therefore uncorrelated X--ray and residual
optical/UV luminosity variations might be expected.

An alternative, more speculative possibility involves coronal activity
driven by shear and convection in the outer disk where matter accumulates.
In this interpretation
there would be virtually no mass accretion towards the BHC in
quiescence (i.e. accumulation in the outer disk takes place without any
significant leakage towards the BHC).

The analogy with the X--ray emission from K stars in RS CVn type binaries
(which emit up to $\sim 10^{32}\ergs$ in the ROSAT band; Dempsey
et al. 1993) suggests that, in short orbital period BHC systems, such as
A 0620--00, the low level X--ray quiescent luminosity
($\sim 10^{31}-10^{32}\ergs$) might also be due to coronal activity of
the companion star (see also Bildsten \& Rutledge 2000).
In order to reach luminosities in excess of $\sim 10^{31}\ergs$ a 
subgiant companion is required (Eracleous et al. 1991).

\subsection{Quiescent neutron star Low Mass X--ray Transients}

In this section we apply the scenario outlined above for BHCs to the case
of NS LMXRTs. We first review the evidence that the NSs in these
systems and, likewise LMXRBs, are fast spinning and weakly magnetic.
We then explore the regimes spanned by such NSs as the mass inflow rate
decreases and argue that radio pulsar shock emission,
together with thermal emission from the NS surface, are responsible for the
the quiescent X--ray emission of NS LMXRTs.

It had long been suspected that the NSs of persistent and
transient LMXRBs have been spun up to very short
rotation  periods by accretion torques (Smarr \& Blandford 1976; Alpar et
al. 1982);  however, conclusive evidence has been obtained only recently.
The most striking case is that of
SAX~J1808.4--3658, a bursting transient source discovered with BeppoSAX in
1996 (in't Zand et al. 1998). In April 1998, RossiXTE observations revealed
a coherent $\sim 401$~Hz modulation, testifying to the presence of magnetic
polar cap accretion onto a fast rotating magnetic NS (Wijnands \& van
der Klis 1998; Chakrabarty \& Morgan 1998).

Millisecond rotation periods have also been inferred for 7 other
LMXRBs of the Atoll (or suspected members of the) group through the
oscillations
that are present for a few seconds during type I X--ray
bursts (for a review see van der Klis 1999).
The spin frequencies inferred from burst oscillations span the
range from 150 to 590~Hz, in agreement with the high spin
frequency of millisecond radio pulsars (of which LMXRBs are likely
progenitors). The $\sim 550$, 525 and 150~Hz signal from Aql~X-1,  KS
1731--260
and the  Rapid Burster, respectively, are the
only burst  oscillations revealed so far from LMXRTs.

Regarding the NS magnetic field, Psaltis \& Chakrabarty (1999) estimate
for SAX~J1808.4--3658 a
value of $B\sim 10^8-10^9$ G, by adopting different models
for the disk-magnetosphere interaction. Indirect evidence for fields of
$B\sim 10^8$ G derives also from the steepening in the X--ray
light curve decay and marked change of the X--ray spectrum
when the luminosity reaches a level of $\sim 10^{36}\ergs$
in Aql X-1 (Campana et al. 1998b; Zhang et al. 1998) and SAX J1808.4--3658
(Gilfanov et al. 1998), once these changes are
interpreted in terms of the onset of the
centrifugal barrier. It is still unclear whether such magnetic fields
are strong enough that a small magnetosphere can survive when
transient and persistent LMXRBs accrete close to their highest rate.

Accretion onto the surface of a magnetic NS can take
place only as long as the centrifugal barrier of the rotating magnetosphere
is open. In this regime, the accretion luminosity is given by $L(R)=G\,M\,
\mdot/R$. Below a critical mass inflow rate $\mdot_{\rm cb}$, corresponding
to
a luminosity of \footnote{We use here simple spherical accretion theory. This
is a reasonably accurate approximation when the accretion
disk at the magnetospheric boundary is dominated by gas pressure, as in NS
LMXRTs for a luminosity of $\leq 10^{36}$~erg~s$^{-1}$. For a more general
approach see e.g. Campana et al. (1998a).}
\be
L_{\rm cb}\simeq 5 \times 10^{35}\, B_8^2\, M_{1.4}^{-2/3}\, R_6^5\,
P_{2.5\,{\rm ms}}^{-7/3}\ {\rm erg\ s}^{-1} \ ,
\label{lmin}
\en
($B=10^8\,B_8$ G, $P=2.5\times 10^{-3}\, P_{\rm 2.5\, ms}$~s, $R=10^6\, R_6
$~cm and $M= 1.4\, M_{1.4}\msole$ are the magnetic field, spin period,
radius
and mass of the NS, respectively)
the magnetosphere expands beyond the corotation radius,
$r_{\rm cor}$, such that the centrifugal barrier closes,
and the matter inflow stops
at the magnetospheric radius, $r_{\rm m}$, therefore releasing a lower
accretion luminosity. The accretion luminosity gap, $\Delta_{\rm c}$,
across the centrifugal barrier is (Corbet 1996):
\be
\Delta_{\rm c} = {{r_{\rm cor}} \over R} = \Bigl( {{G\,M\,P^2}\over
{4\,\pi^2\,R^3}} \Bigr)^{1/3}
\simeq 3\, P_{\rm 2.5\, ms}^{2/3}\,M_{1.4}^{1/3}\,R_6^{-1} \ .
\label{gapa}
\en
$\Delta_{\rm c}$ depends almost exclusively on the
spin period $P$ and is basically a measurement of how deep $r_{\rm cor}$ is
in the potential well of the NS.

Once the centrifugal barrier is closed, the
NS enters  the so-called propeller stage.
It should be noted that the luminosity released by accretion in the
propeller regime ($L(r_{\rm m})=G\,M\,\mdot/ r_{\rm m}$), is only a lower
limit.
An additional luminosity might be produced by: (a) matter leaking through the
barrier (especially from the higher latitudes) and accreting onto the
NS surface (Stella et al. 1986; Corbet 1996; Zhang et al. 1998);
(b) the NS rotational energy released in the
disk through the interaction with the magnetic field of the NS
(e.g. Priedhorsky 1986).
Moreover the fate of the bulk of the inflowing matter is uncertain as it may
accumulate or be expelled by the action of the supersonically rotating
magnetospheric boundary (Davies \& Pringle 1981).

As the mass inflow rate decreases further the magnetosphere expands
until the light cylinder radius, $r_{\rm lc}=c\,P/2\,\pi$, is reached;
beyond this point the radio pulsar dipole radiation will turn on and
begin pushing outward the inflowing matter, due
to a flatter radial dependence of its pressure compared to that of
disk or radial accretion inflows
(Illianorov \& Sunyaev 1975; Stella et al. 1994; Campana et al. 1995).
The equality $r_{\rm m}=r_{\rm lc}$ therefore defines the lowest mass
inflow rate
(and therefore accretion luminosity) in the propeller regime.
An accretion luminosity ratio of
\be
\Delta_{\rm p}=\Bigl( {{r_{\rm lc}}\over {r_{\rm cor}}} \Bigr)^{9/2} \simeq
440\, P_{\rm 2.5\, ms}^{3/2}\,M_{1.4}^{-3/2}
\label{gapp}
\en
characterises the range over which the propeller regime applies.
Note that also this ratio depends mainly on the spin period $P$.
Once in the rotation powered regime, a fraction $\eta$
of the spin down luminosity, $L_{\rm sd}$, converts to shock emission
in the interaction between the relativistic wind of the
NS and the companion's matter flowing through the Lagrangian point.
Theoretical models indicate that the material lost by the
companion star may take somewhat different shapes, ranging from a bow shock
to an irregular annular region in the Roche lobe of the NS,
depending on radio pulsar wind properties and
the rate and angular momentum of the mass loss from the companion star
(Tavani \& Brookshaw 1993; note that if a bow shock forms, which prevents
material from accumulating in the Roche lobe of the compact object, the
outburst activity might be inhibited permanently, see e.g. Shaham \&
Tavani 1991). The shock luminosity can be expressed as $L_{\rm shock}
=\eta\,L_{\rm sd}\sim 6\times 10^{31}\,\eta_{-1}\,B^2_8\,P_{-2}^{4}\ergs$
with $\eta$ as large as 0.1 (Tavani 1991; $\eta\sim \eta_{-1}\,0.1$).
The luminosity ratio across the transition from the propeller to the
rotation powered regime can be approximated as (Stella et al. 1994;
Campana et al. 1998a)
\be
\Delta_{\rm s}={ 3 \over{2\,\sqrt{2}\,\eta}} \, \Bigl( {{r_{\rm g}}
\over {r_{\rm lc}}}\Bigr)^{1/2}\simeq 2\,\eta_{-1}^{-1}\,
P_{\rm 2.5 \, ms}^{-1/2}\,M_{1.4}^{1/2} \ ,
\label{gaps}
\en
\noindent where $r_{\rm g}=G\,M/c^2$ is the gravitational radius.
Currently no prescription is available
for the dependence of the shock-powered luminosity on the mass inflow
rate. We note, however, that the 0.5--10 keV shock luminosity of
the radio pulsar/Be star binary PSR 1259--63 varied
by a factor of  $\sim 10$ in response to the $\gsim 3$ orders of magnitude
variation of mass capture rate at the accretion radius
across its highly eccentric orbit
(Hirayama et al. 1999). This suggests that the shock-powered luminosity
depends only very weakly on mass inflow rate and, also in consideration of
the uncertainties in the value of $\eta$, we will consider it a constant.
The energy spectrum due to shock emission is expected to be a power law with
photon index of $\sim 1.5-2$ that extends from a $\sim 10$ eV to $\sim 100$
keV, with both energy boundaries shifting as $B_8\,P_{\rm 2.5 ms}^{-3}$
(Tavani \& Arons 1997; Campana et al. 1998a).

In summary, as the mass inflow rate towards a fast rotating weakly magnetic
NS decreases, the ratio of the minimum luminosity from accretion onto the NS,
$L_{\rm cb}$,
(just before the onset of the propeller effect) to the shock luminosity
in the rotation powered regime is
\begin{eqnarray}
\Delta_{\rm cps} & \equiv & \Delta_{\rm c} \,\Delta_{\rm p}\, \Delta_{\rm s}
 \nonumber \\
& = & {3 \over{2\,\sqrt{2}\,\eta}}\,r_{\rm lc}^4\,r_{\rm cor}^{-7/2}\,
r_{\rm g}^{1/2}\,R^{-1} \nonumber \\
& \simeq &
2\times 10^3 \,\eta_{-1}^{-1}\,P_{\rm 2.5\, ms}^{5/3}\,
M_{1.4}^{-2/3} R_6^{-1}\ .
\label{gapap}
\end{eqnarray}
As expected $\Delta_{\rm cps}$ depends almost exclusively on the spin
period.
On the contrary the absolute luminosity scale, set e.g. by $L_{\rm cb}$,
depends also on the NS magnetic field (see Eq.~1).
It is easy to show that the variation of the mass
inflow rate corresponding to $\Delta_{\rm cps}$ can be expressed as
\be
\Delta\dot M_{\rm cps} \simeq 1\times 10^2\,P_{\rm 2.5\, ms}^{7/6}\,
M_{1.4}^{-7/6}\ .
\en

By combining Eqs.~1 and 5 with the Eddington luminosity, a lower limit is
obtained on the minimum to
maximum X--ray luminosity ratio of NS LMXRTs within the model discussed
here; this is (cf. Fig. 2)
\begin{eqnarray}
{{L_{\rm min}}\over{L_{\rm max}}} &>& {{L_{\rm cb}}\over{L_{\rm Edd}\,
\Delta_{\rm cps}}} = {{\eta\,L_{\rm sd}}\over{L_{\rm Edd}}} \nonumber \\
&\sim& 2\times 10^{-6}\,\eta_{-1}\,B_8^2\,M_{1.4}^{-1}
\,R_6^7\,P_{\rm 2.5ms}^4 \ .
\end{eqnarray}
For the spin period and magnetic field inferred for NS LMXRTs this limit
is consistent with the range of observed minimum to maximum X--ray
luminosity ratios discussed in Section 2.2. On the contrary
the minimum to maximum X--ray luminosity ratio inferred for BHC systems
is substantially lower (see Section 2.1). The corresponding
(minimum) mass inflow rate variation is given by
\be
{{\mdot_{\rm cb}}\over{\mdot_{\rm Edd}\,\Delta \mdot_{\rm cps}}}
\sim 3\times 10^{-5}\,B_8^2\,M_{1.4}^{-11/6}\,R_6^{5}
\,P_{\rm 2.5ms}^{-7/2} \ .
\en
A 7--8 orders of magnitude variation of mass inflow rate from
outburst to quiescence (see Section 4.1) would easily encompass this range.
It should be noticed that two additional contributions to the
quiescent X--ray luminosity of NS LMXRTs might be expected, the X--ray
spectrum
of which is thermal-like. An X--ray component in the
$10^{32}-10^{33}\ergs$ range from the whole NS surface should be produced by
the release of thermal energy from the NS interior heated up during the
accretion episodes, unless a pion condensate is present in the NS core
(Campana et al. 1998a; Brown, Bildsten \& Rutledge 1998; Rutledge et al.
1999).
Note that this might be relevant to both the propeller and the rotation
powered pulsar regimes.
Heating by relativistic particles associated with the radio pulsar
emission might cause an additional soft X--ray quasi-thermal component
from a much reduced area in the vicinity of the NS magnetic poles.
The latter component is expected to be pulsed. The analogy with X--ray
properties of the millisecond radio pulsars
PSR J0437--4715 ($P=5.8$~ms, $B=3\times10^8$~G) and PSR J0218+4232
($P=2.3$~ms, $B=4\times10^8$~G) suggests an X--ray luminosity from the
polar caps in the $10^{30}-10^{31}\ergs$ range (Becker \& Tr\"umper 1999).
The possible role of these emission components is further discussed in
Section 5.

\begin{figure*}[!htb]
\psfig{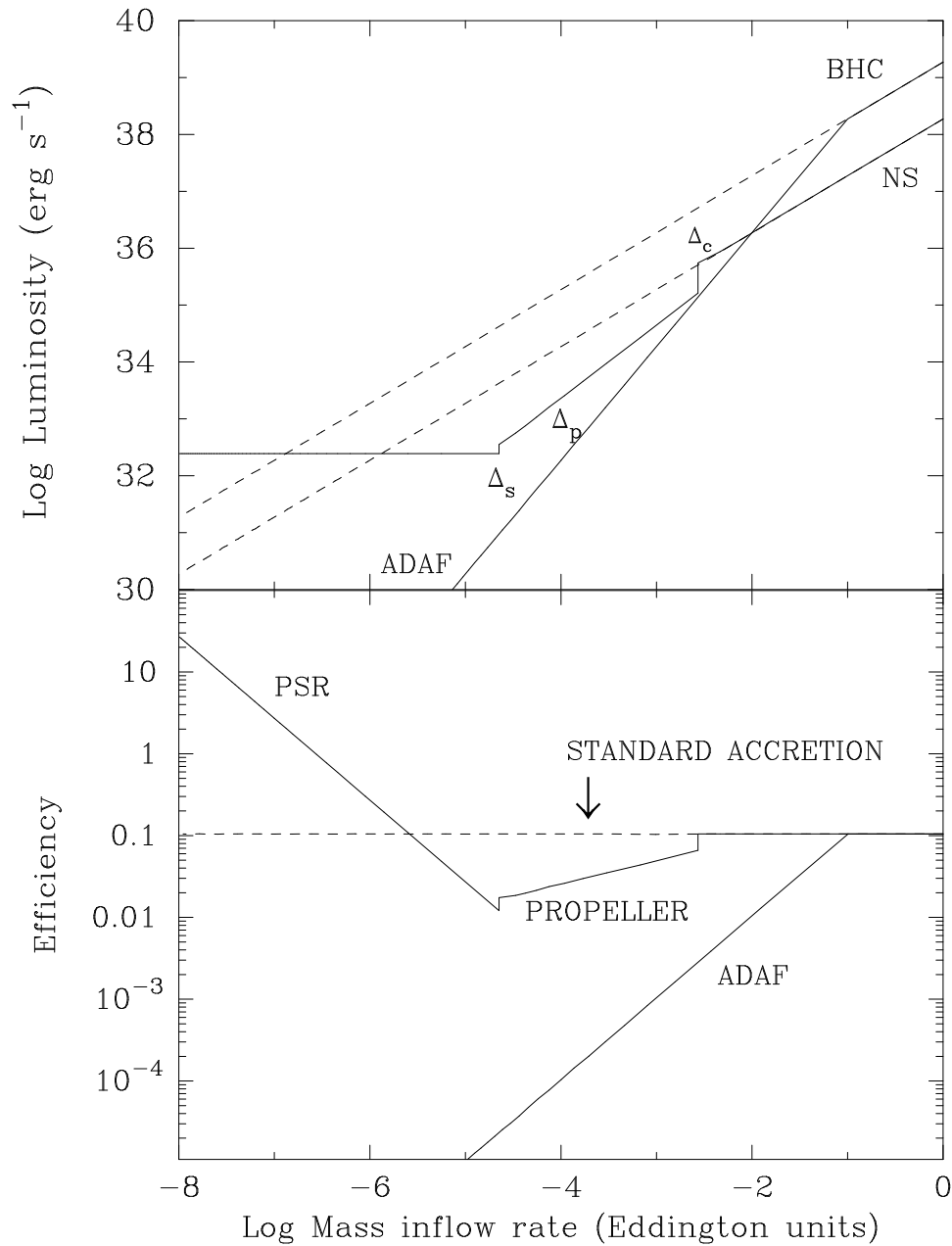}
\caption[h]{The upper panel shows the luminosity versus
the mass inflow rate (in Eddington units
$\mdot \sim 1.4\times 10^{18}\,M/M_{\odot}$ g s$^{-1}$)
for different accretion regimes onto BHs and NSs.
The upper line refers to a $14\msole$ BH.
The dashed line marks the standard accretion regime
(efficiency $\epsilon = 0.1$) and
the continuous line the ADAF model. The lower lines refer to accretion
onto a $1.4\msole$ NS. The dashed line refers to accretion onto the NS
surface. The continuous line gives the luminosity produced by
a 2.5 ms spinning, $B=10^8$ G NS in different regimes.
The transition from the standard accretion regime ($L \propto \mdot$)
to the regime of accretion onto the magnetospheric
boundary in the propeller regime ($L \propto \mdot^{9/7}$; cf. Stella
et al. 1994) and,
subsequently,  the radio pulsar shock emission regime ($L \sim const$)
is clearly seen for decreasing mass inflow rates (see text).
The lower panel gives the mass inflow to radiation conversion efficiency
in the different regimes.}
\end{figure*}

Fig. 2 illustrates the different emission regimes discussed in the previous
sections for NS and BHC in LMXRTs as a function of the mass inflow rate.
Accounting for the measured X--ray luminosity of quiescent BHCs (see Table 1)
by means of standard ($\epsilon\sim 0.1$) accretion
would require a very low accretion rate of
$10^{-8}-10^{-6}\,\mdot_{\rm Edd}$.
Any substantial X--ray flux from, e.g., coronal activity in
the outermost disk regions would correspondingly decrease the required
quiescent mass accretion rate into the BHC.

In the case of a NS system the different regimes are clearly seen for
decreasing mass inflow rates: $L \propto \mdot$ when accretion onto the NS
surface takes place;
$L \propto \mdot^{9/7}$ from accretion onto the magnetospheric
boundary in the propeller regime; $L \sim const$ in the radio
pulsar shock emission regime. These regimes are separated
by the gaps characterising the onset of the propeller and radio pulsar
regimes.
It is apparent from Fig.~2 that for a quiescent mass inflow rate
of $\lsim 10^{-5}\,\mdot_{\rm Edd}$, the shock emission radio pulsar
regime applies (cf. Eq. 8).
Correspondingly, an X--ray luminosity of $\eta\, L_{\rm sd} \sim
2\times 10^{32}\,\eta_{-1}\,B_8^2\,P_{2.5\,{\rm ms}}^{-4}\ergs$ would be
expected, which is in the range measured from quiescent LMXRTs (cf. Eq. 7).
It is also apparent from Fig. 2 that the X--ray luminosity swing predicted
by our model is 1--2.5 decades smaller in NS than in BHC systems,
if comparably large swings of mass inflow rates (in Eddington units)
characterise the two classes of LMXRTs. This is also in agreement with the
observations (see Section 2).

Incidentally, we note that in the relatively slow ($P\geq 1$~s)
and high magnetic field ($B\sim 10^{12}$~G) NSs that are usually
found in Be star X--ray transients, the centrifugal barrier sets in around
$L_{\rm cb}\sim 10^{35}-10^{36}\ergs$, $\Delta_{\rm c}$ is a few
hundreds and the propeller regime applies down to very low
mass inflow rates ($\sim 10^{-6}-10^{-7} \dot M_{\rm Edd}$).
For a plausible swing
of mass inflow rate across the outburst to quiescence transition of Be
transient systems, the resulting accretion luminosity swing
is therefore expected to be many orders of
magnitude larger than standard accretion theory would predict, without
the NS ever entering the radio pulsar regime.

\section{Discussion}

As emphasised in Section 4.2, the shock emission spectrum
in the radio pulsar regime is expected to be
a power law with photon index of $\Gamma \sim 1.5-2$. This is, at least
qualitatively, in agreement with the hard power law like X--ray component
observed in the quiescent X--ray spectrum of Cen~X-4, Aql X-1 and
X~1732--304.
The extended power law spectrum expected from shock emission
is also in agreement with the recently determined residual UV spectrum
of Cen~X-4, which shows no evidence for a turnover down to lowest measured
UV energies ($\sim 7.5$~eV)
and matches quite well the extrapolation of the (power law) X--ray spectrum.
We note that if the shock is located close to the circularisation
radius (the most conservative case), an accretion luminosity in the
$\sim 10^{32}\ergs$ range would be released by the stream of matter from
the companion, with a (black body) temperature of
7000--10000 K. This is, in turn, consistent with the idea that
shock emission is at least comparable with the emitted spectrum in
the blue and UV band, as expected in our model for typical NS parameters
of LMXRTs.
On the contrary quiescent state observations of the BHC LMXRT A~0620--00
show a marked steepening of the UV spectrum above $\sim 5$ eV, while the
X--ray luminosity is one order of magnitude lower than that of Cen~X-4.

Concerning the soft X--ray thermal-like component
observed in the NS LMXRTs Aql~X-1 and Cen~X-4
(see Section 2.2), this accounts for about $\sim 60\%$ and $\sim 55$\% of
the total 0.5--10 keV quiescent luminosity, respectively. Therefore, the
conclusions in Section 4.2 concerning the matching of the observed
quiescent luminosity with the luminosity of the shock emission
(power law) component
remain essentially unchanged. We note that the effective black body radii
inferred from spectral fitting of the soft component
($\sim 0.8$ and $\sim 3.1$~km, respectively) are substantially smaller
than the NS radius.
However detailed, radiative transfer calculations for the NS
atmosphere indicate that the emergent thermal-like X--ray spectrum is
complex and simple black body fits are likely to underestimate the
effective emission radius and overestimate the temperature
by a factor of 3--10 and 2--3, respectively (Rutledge et al. 1999).
Consequently, it cannot be ruled out yet that thermal emission
from the whole NS surface powers the soft X--ray component of
quiescent NS LMXRTs.
Alternatively heating of the magnetic polar caps
by relativistic electrons might be
responsible for the soft X--ray component observed from NS LMXRTs.

One might think of a propeller model in which the soft thermal-like
component of the quiescent X--ray spectrum is indeed produced by cooling
of the NS surface. In this case the requirement on the rate at which
the inflowing matter reaches the NS magnetic pole could be relaxed
(see M99), such that virtually 100\% of the inflowing matter can
be halted at the magnetospheric boundary.
If we adopt a quiescent mass inflow rate that
is $\sim 1/3$ of the binary mass transfer $\mdot_{\rm tr}$
(in analogy with the values deduced from ADAF modeling of
quiescent BHCs, see M99), it is easy to see that for the values
of $\dot M_{\rm tr}/3 \sim 10^{15}-10^{16} \gs$
($10^{16}\gs$ pertains to the longest orbital period BHCs)
that are expected for NS
systems, the luminosity produced in propeller regime by accretion onto
the magnetosphere would be only a few times less efficient than
accretion onto the NS surface, giving rise to a
2--3 orders of magnitude larger luminosity than observed (see Fig. 2).
Such a model could be made marginally viable, only if accretion onto the
magnetospheric boundary took a place in quiescent NS LMXRTs at rate
a factor of $\sim 100$ smaller than $\dot M_{\rm tr}/3$ (note that
the relevant values of the mass inflow rate $\sim 10^{13}-10^{14}\gs$
would straddle the boundary between the propeller and the
radio pulsar regime). However, this would require abandoning the idea
that the swing of mass inflow rates is similar across BHC and NS systems.

We remark that if the radio pulsar/shock emission regime applies to
quiescent NS systems, accretion towards the NS can resume only after
the pressure of the accumulating material overcomes the radio pulsar
pressure. This translates to a condition on the recurrence time
of the outbursts
\be
\Delta t \gsim 0.3\,\mdot_{15}^{-1}\,h_{9}\,T_4^{-1}\,B_8^2\,P_{\rm 2.5\,ms}^{-4}\
{\rm yr}
\en
where $\mdot_{15}$ is the mass transfer rate from the companion
in units of $10^{15}\gs$, $h_9$ is the height of the outer disk in which
matter accumulates in units of $10^9$ cm and $T_4$ its temperature
in units of $10^4$ K.
We note that the inferred outburst recurrence times of NS LMXRTs
are indeed longer than $\Delta t$ in the expression above for
typical parameters.
This, in turn, suggests that once the instability sets in,
the radio pulsar pressure would be unable to halt the
inflowing matter, and accretion can resume unimpeded.

\section{Conclusions}

The scenario we explored in this paper for the quiescent emission
of LMXRTs assumes only very little
matter (if at all) proceeds toward the collapsed object
while most of the mass transferred from the companion star accumulates
in an outer disk region or is lost in a wind.
This idea is in line with models designed to explain the
delays between the optical and UV light curves in the outbursts of
cataclysmic variables. A suppression of the innermost disk regions
(or nearly so) during quiescence is envisaged in those models as well
(e.g. Livio \& Pringle 1992; Meyer \& Meyer-Hofmeister 1994; King 1997).
Our suggested scenario is also reminiscent of earlier (pre-ADAF) 
disk instability
models that were proposed  to explain the long recurrence time of A 0620--00
within the context of standard accretion theory (Huang \&
Wheeler 1989; Mineshige \& Wheeler 1989). 
At present we can only speculate on the reason why
the quiescent outer disk would remain stable despite its
effective temperature in $\sim 5000-10000$~K range. Perhaps the {\it ad hoc}
variation of the $\alpha$ viscosity parameter that is required to
make current disk instability models reproduce
the outburst recurrence times (see Lasota \& Hameury 1998) does not take
place at the temperature at which the hydrogen ionisation and opacity
change (as in current models), but rather at
somewhat higher temperatures. In this case the outer disk region of
quiescent LMXRTs could be in the lower branch of the corresponding
surface density vs. viscosity relationship.

In our model the residual quiescent optical/UV emission (after subtraction
of the contribution from the mass donor star) in BHC systems derives
entirely from the gravitational energy released by the matter transferred
from the companion star at radii comparable to the circularisation radius.
The low quiescent X--ray luminosity originates either from standard
accretion into the BHC at very low rates (some $\sim 10^{11}-10^{12}\gs$),
and/or from coronal activity of the
outer disk where matter accumulates or, limited to  short orbital
period systems, the companion star which is forced to corotate (see
also Bildsten \& Rutledge 2000).
For comparably small mass inflow rates ``leaking" from the outer disk
regions, it can be safely concluded that the NSs of LMXRTs are in the
radio pulsar regime, if their spin periods are a few millisecond
and magnetic field some $\sim 10^8-10^9$~G, as indicated by the
observations. The interaction of the radio pulsar relativistic wind with
the matter transferred from the companion star
gives rise to a shock, the power law like emission of which powers the
harder X--ray emission component and optical/UV excess which are observed
at a level of $\sim 10^{32}-10^{33}\ergs$ in the quiescent state of Cen~X-4.
The soft, thermal-like component which contributes about half of
the quiescent X--ray luminosity of several NS LMXRTs arises from the
cooling of the NS surface in between outbursts or, perhaps,
heating of the magnetic polar caps by relativistic particles in the radio
pulsar magnetosphere.

Both the X--ray and bolometric luminosity swing of
of BHC as well as NS systems are well matched by the model,
for comparable ratios of minimum to
maximum mass inflow rates toward the collapsed object
across the two classes of LMXRTs.
Moreover, different predictions are made about the quiescent emission
of LMXRTs, which could be tested through
higher spectral resolution and throughput observations to be obtained
in the near future (e.g. with Newton-XMM). For example, if the
quiescent X--ray emission of BHC systems resulted from coronal activity,
emission lines from heavy elements and an optically thin thermal
spectrum would be expected. Moreover residual optical/UV and X--ray flux
variations should be correlated. On the contrary if a hot accretion disk
with standard (as opposed to ADAF) efficiency gives rise to the X--ray flux,
uncorrelated X--ray and residual optical/UV variations might take place.
Variations in the optical/UV excess and the X--ray power law component
of quiescent NS systems should be correlated, if they both arose from
radio pulsar shock emission. The detection of
pulsations at the NSs spin in the quiescent soft X--ray component
would rule out emission from the whole NS
surface and argue in favor of heated magnetic polar caps.
The ultimate test of the radio pulsar regime in quiescent NS LMXRTs
would be the detection of a pulsed radio signal. Yet, the matter in the
outer disk and/or the shock might enshroud the pulsar making any radio signal
very difficult to detect (Kochanek 1993; Stella et al. 1994; Campana et al.
1998a). The geometry of the radio pulsar
shock and matter accumulating during the quiescent intervals
of NS LMXRTs is highly uncertain; detailed Balmer line Doppler mapping
(possibly at different times after the end of an outburst) could provide
important clues on this issue.

\begin{acknowledgements}
This work was partially supported through ASI grants. LS acknowledges
useful discussions with Luciano Burderi and Phil Charles. Paolo
Goldoni provided useful comments on an earlier version of this manuscript.

\end{acknowledgements}

\end{document}